\documentclass[aps,12pt,final,notitlepage,onecolumn,oneside,nobibnotes,
nofootinbib,superscriptaddress,centertags,showpacs]{revtex4}

\usepackage{amsfonts}

 \def\1{\'{\i}}                           
\def\k{\kappa}                           
\def\Sk{{\rm\ \!S}}            
\def\Ck{{\rm\ \!C}}           
\def\Tk{{\rm\ \!T}}

\def\be{\begin{equation}}
\def\ee{\end{equation}}
\def\bea{\begin{eqnarray}}
\def\eea{\end{eqnarray}}

\def\dd{{\rm d}}
\def\>#1{{\mathbf#1}}                 
\def\te{\phi}

\def\tea{r}

\begin{document}

 \title{Superintegrability on N-dimensional  
spaces of constant curvature from $so(N+1)$ and its contractions\footnote{Based on the contribution presented at the ``XII International Conference on Symmetry Methods in Physics",   Yerevan (Armenia),  July  2006.\\
To appear in Physics of Atomic Nuclei.}}
\author{\firstname{Francisco J.}~\surname{Herranz}}
\email{fjherranz@ubu.es}
\affiliation{Departamento de F\1sica, Escuela Polit\'ecnica Superior,
Universidad de Burgos, E-09006, Burgos, Spain.}
\author{\firstname{\'Angel}~\surname{Ballesteros}}
\email{angelb@ubu.es}
\affiliation{Departamento de F\1sica, Facultad de Ciencias,  Universidad de Burgos,
E-09006, Burgos, Spain.}

\begin{abstract}
\baselineskip=16pt

\noindent
The Lie--Poisson algebra $so(N+1)$ and some of its   contractions  are used to construct
a family of superintegrable Hamiltonians on   the $N$D spherical, Euclidean, hyperbolic,
Minkowskian and (anti-)de Sitter spaces.  We firstly present a Hamiltonian which is a
superposition of   an arbitrary central potential with $N$ arbitrary centrifugal terms. Such a
system is quasi-maximally superintegrable since this is  endowed with $2N-3$ functionally
independent constants of the motion (plus  the Hamiltonian). Secondly,  we identify two
maximally superintegrable Hamiltonians  by choosing a specific central potential and finding at
the same time the remaining integral.  The former is the generalization of  the
Smorodinsky--Winternitz system to the above six spaces, while the latter is a generalization
of  the Kepler--Coulomb potential, for which  the Laplace--Runge--Lenz $N$-vector   is also
given.  All the systems and constants of the motion are explicitly expressed   in a unified
form in terms of ambient and polar coordinates as they are parametrized  by two contraction
parameters  (curvature  and signature of the metric).
\end{abstract}



\maketitle

\newpage

\baselineskip=17pt


\section{Introduction}

\noindent
Let us consider the following potential on the $N$-dimensional ($N$D) Euclidean space~\cite{6}:
\be
{\cal U}={\cal F}(r)+\  \sum_{i=1}^N \frac{\beta_i}{x_i^2},
\label{aa}
\ee
where $\beta_i$ are arbitrary real constants,  $x_i$ are Cartesian coordinates and ${\cal
F}(r)$ is an arbitrary smooth function depending on the Euclidean distance $r
=\left(\sum_{i=1}^N x_i^2\right)^{1/2}$.   This potential is is known to be superintegrable
and can be interpreted as the superposition of a central term with $N$ centrifugal barriers
associated to the  $\beta_i$'s. Furthermore, two particular choices of ${\cal
F}(r)$ provide  well known maximally superintegrable (MS) Euclidean Hamiltonians: 
\begin{itemize}
\itemsep=0pt
\item  If  ${\cal F}(r)= \omega^2 r^2$ we obtain an
isotropic harmonic oscillator with angular frequency $\omega$.  The $N$ arbitrary
centrifugal terms can be added to the oscillator potential keeping maximal superintegrability;
the resulting potential is the Smorodinsky--Winternitz (SW) system~\cite{9,7,8,10}.

\item If  ${\cal
F}(r)= -k/r$, where $k$  is a real constant, we find the Kepler--Coulomb (KC) potential. In
this case, only a maximum number of $(N-1)$   centrifugal terms can be considered in
order to preserve  maximal superintegrability~\cite{6,Miguel,Williams}.

\end{itemize}

Our aim in this paper is  to present a unified generalization of the   superintegrable
potential (\ref{aa}) and its two particular MS  cases to the $N$D  spherical,
Euclidean, hyperbolic, Minkowskian and both de Sitter spaces. The approach we shall make use is
based on the Lie--Poisson algebras associated to the Lie groups and subgroups involved in the
construction of the above spaces as symmetrical homogeneous ones. Thus in the next section we
introduce
 the   basics on the  Lie groups of
isometries on these six $N$D spaces together with the two coordinate systems we shall deal
with:   $(N+1)$ ambient  coordinates in an auxiliary linear  space   and  $N$
intrinsic geodesic polar (spherical) coordinates. The kinetic energy which gives rise to the
geodesic motion is   studied in section III by starting from the metric. 
 The   generalization  of the potential  (\ref{aa}) is
addressed in section IV in such a manner that
general and global expressions for the Hamiltonian and  its
$2N-3$ functionally independent integrals of motion are explicitly given. Finally,  last section
is   devoted to the study of the two MS Hamiltonians arising in the above
family by choosing in an adequate way the radial function  and by finding at the same time the
remaining constant of the motion. Consequently, we obtain the generalization of the
SW and generalized   KC   potentials   for any value of the
curvature and signature of the metric. We remark that these results generalized to arbitrary
dimension $N$ the 3D case recently studied in~\cite{kiev}.


\section{Riemannian spaces and relativistic spacetimes}
\noindent

Let us consider a set of real Lie algebras $so_{\k_1,\k_2}(N+1)$ which come from    $\mathbb
Z_2^{\otimes N}$
  graded contractions of $so(N+1)$,   where
$\k_1$ and $\k_2$ are two real contraction parameters. The non-vanishing  Lie brackets of
$so_{\k_1,\k_2}(N+1)$ in the basis spanned by
$\{J_{\mu\nu}\}$ ($\mu,\nu=0,1,\dots,N$; $\mu<\nu$) read~\cite{CKND}:
\begin{equation}
\begin{array}{lll}
[J_{ij},J_{ik}]=  J_{jk},&\quad [J_{ij},J_{jk}]=-J_{ik},&\quad [J_{ik},J_{jk}]=J_{ij},\\[2pt]
[J_{1j},J_{1k}]= \k_2 J_{jk},&\quad [J_{1j},J_{jk}]=-J_{1k},&\quad
[J_{1k},J_{jk}]=J_{1j},\\[2pt] 
[J_{01},J_{0k}]= \k_1 J_{1k},&\quad [J_{01},J_{1k}]= 
-J_{0k},&\quad [J_{0k},J_{1k}]=\k_2 J_{01},\\[2pt]
 [J_{0j},J_{0k}]= \k_1\k_2 J_{jk},&\quad [J_{0j},J_{jk}]= -   J_{0k},&\quad
[J_{0k},J_{jk}]=  J_{0j},
\end{array}
\label{aax}
\end{equation}
where $i,j,k=2,\dots,N$ and $i<j<k$. Both contraction parameters $\k_l$ $(l=1,2)$ can take any
real value.   By scaling the Lie generators, each $\k_l$ can be reduced to either
$+1$, 0 or $-1$; the limit $\k_l\to 0$ is   equivalent to apply an In\"on\"u--Wigner
contraction.

The quadratic Casimir for $so_{\k_1,\k_2}(N+1)$, associated to the Killing--Cartan
form, is given by
\begin{equation}
{\cal C}= \k_2
J_{01}^2+\sum_{j=2}^N J_{0j}^2 +\k_1 \sum_{j=2}^N J_{1j}^2 +\k_1\k_2\sum_{i,j=2}^N J_{ij}^2 .
\label{bba}
\end{equation}

Next from the Lie group  $SO_{\k_1,\k_2}(N+1)$  with Lie algebra (\ref{aax}) we construct the
following
$N$D symmetrical homogeneous space: 
\be
{\mathbb
S}^N_{[\k_1]\k_2}=SO_{\k_1,\k_2}(N+1)/SO_{\k_2}(N),\qquad
SO_{\k_2}(N)=\langle J_{ij}; i,j=1,\dots,N\rangle .
\label{bca}
\ee
The parameter $\k_1$ turns out to be the constant sectional curvature of the space, while $\k_2$
determines the signature of the metric   as ${\rm diag}(+1,\k_2,\dots,\k_2)$. In this way, we
find that ${\mathbb S}^N_{[\k_1]\k_2}$ comprises well known spaces of constant curvature:

\begin{itemize}
\itemsep=0pt
\item  When $\k_2$ is positive, say $\k_2=+1$, we recover the
three classical Riemannian spaces. These are the spherical $\k_1>0$, Euclidean $\k_1=0$, and
hyperbolic spaces $\k_1<0$:
${\mathbb S}^N_{[+]+}=SO(N+1)/SO(N)$, ${\mathbb S}^N_{[0]+}=ISO(N)/SO(N)$
and ${\mathbb S}^N_{[-]+}=SO(N,1)/SO(N)$.
 The first two rows of (\ref{aax}) span the rotation Lie subalgebra
$so(N)$, while the $N$ generators $J_{0i}$ $(i=1,\dots,N)$ appearing in the last two rows play
the role of translations. The curvature can be written as $\k_1=\pm 1/R^2$ where $R$ is the
radius of the space  ($R\to \infty$ for the Euclidean case).

\item  When $\k_2$ is negative we find a Lorentzian metric corresponding to  
 relativistic spacetimes; namely, the anti-de Sitter $\k_1>0$, Minkowskian $\k_1=0$, and
de Sitter spaces $\k_1<0$:
${\mathbb S}^N_{[+]-}=SO(N-1,2)/SO(N-1,1)$, ${\mathbb S}^N_{[0]-}=ISO(N-1,1)/SO(N-1,1)$ and
${\mathbb S}^N_{[-]-}=SO(N,1)/SO(N-1,1)$. The generators $J_{01}$, $J_{0j}$, $J_{1j}$ and
$J_{ij}$ $(i,j=2,\dots,N)$ are identified with time
translation, space translations, boosts and spatial rotations,
respectively. The first row of (\ref{aax}) is a rotation subalgebra $so(N-1)$ and the two first
rows span the Lorentz subalgebra   $so(N-1,1)$. The two contraction parameters can be expressed
as  $\k_1=\pm 1/\tau^2$, where $\tau$ is the (time) universe
radius, and $\k_2=-1/c^2$, where $c$ is the speed of light.

\item  Finally, the contraction $\k_2=0$ gives rise to Newtonian (non-relativistic) spacetimes
with a degenerate metric. Since we shall construct
superintegrable systems on ${\mathbb
S}^N_{[\k_1]\k_2}$, for which the kinetic energy is provided
by the metric, hereafter we assume $\k_2\ne 0$.

\end{itemize}


In what follows we introduce an explicit model of the space ${\mathbb
S}^N_{[\k_1]\k_2}$ in terms of $(N+1)$ ambient coordinates and also of $N$ intrinsic
geodesic quantities.

The  vector representation  of ${so}_{\k_1,\k_2}(N+1)$  is given
by  $(N+1)\times (N+1)$ real matrices~\cite{CKND}:
\be 
\begin{array}{l}
J_{01}=-\k_1 e_{01}+e_{10},\qquad
J_{0j}=-\k_1\k_2 e_{0j}+e_{j0},\\
J_{1j}=- \k_2 e_{1j}+e_{j1},\qquad J_{jk}=-   e_{jk}+e_{kj},\qquad j,k=2,\dots,N, 
\end{array}
\label{add}
\ee
where $e_{ij}$ is the matrix with   entries
$(e_{ij})_m^l=\delta_i^l\delta_j^m$. Any   generator $X\in {so}_{\k_1,\k_2}(N+1)$  fulfils 
\be
X^T \mathbb I_{\k}+\mathbb I_{\k} X=0 ,
\qquad \mathbb I_{\k}={\rm diag}(+1,\k_1,\k_1\k_2,\dots,\k_1\k_2) ,
\label{af}
\ee
so that    any element $G\in 
SO_{\k_1,\k_2}(N+1)$ verifies $G^T \mathbb I_{\k}   G=\mathbb I_{\k} $.   Then
$SO_{\k_1,\k_2}(N+1)$ is a group of isometries of  $\mathbb I_{\k}$ acting on
a linear ambient space
$\mathbb R^{N+1}=(x_0,x_1,\dots,x_N)$ through matrix multiplication. The origin $\cal O$ in ${\mathbb
S}^{N}_{[\k_1]\k_2}$ has ambient coordinates ${\cal O} =(1,0,\dots,0)$ which is invariant under
the (Lorentz) rotation subgroup $SO_{\k_2}(N)$ (\ref{bca}). The orbit
of ${\cal O}$ corresponds to the   homogeneous space ${\mathbb S}^{N}_{[\k_1]\k_2}$ which is
contained in the ``sphere" provided by $\mathbb I_{\k}$:
\begin{equation}
\Sigma\equiv x_0^2+\k_1 x_1^2+\k_1\k_2 \sum_{j=2}^N x_j^2  =1 .
\label{cd}
\end{equation}

 The $(N+1)$ {\em ambient coordinates}  $\>x= (x_0,x_1,\dots,x_N)$, subjected to
(\ref{cd}),  are also called {\it Weierstrass coordinates}.  The metric on ${\mathbb
S}^N_{[\k_1]\k_2}$
follows from    the flat ambient metric in $\mathbb R^{N+1}$ in the form:
\begin{equation}
{\rm d} s^2=\left.\frac {1}{\k_1}
\left({\rm d} x_0^2+   \k_1 {\rm d} x_1^2+\k_1\k_2  \sum_{j=2}^N  {\rm d} x_j^2 
\right)\right|_{\Sigma}.
\label{ce}
\end{equation}
A differential realization of $so_{\k_1,\k_2}(N+1)$  (\ref{aax}),  coming directly from the
vector representation (\ref{add}), is given by  
\be 
\begin{array}{l}
J_{01}=\k_1 x_1\partial_0 -x_0\partial_1,\qquad
J_{0j}=\k_1\k_2 x_j\partial_0 -x_0\partial_j,\\
J_{1j}=\k_2 x_j\partial_1 -x_1\partial_j,\qquad J_{jk}= x_k\partial_j -x_j\partial_k,
\end{array}
\label{cf}
\ee
 where $ j,k=2,\dots,N$ and  $\partial_\mu=\partial/\partial x_\mu$.

  Next we parametrize the $(N+1)$ ambient coordinates $\>x$ of a generic point ${\cal P}$ in
terms of $N$ intrinsic quantities
$(\tea,\theta,\te_3,\dots,\te_N)$ on  the space
 ${\mathbb S}^N_{[\k_1]\k_2}$   through the following action of $N$
one-parametric subgroups of $SO_\k(N+1)$ on the origin $\cal O$:
\be
   \>x =  \exp(\te_N J_{N-1\, N})\exp(\te_{N-1} J_{N-2\, N-1})\dots\exp(\te_3
J_{23})\,\exp(\theta
J_{12})\,
\exp(\tea J_{01})\, \cal O . 
\label{ba}
\ee
This gives $(i=2,\dots,N-1)$:
 \be 
\begin{array}{l}
x_0 
= \Ck_{\k_1}(\tea),\\
x_1=  \Sk_{\k_1}(\tea)\Ck_{\k_2}(\theta ),\\
x_i 
=\Sk_{\k_1}(\tea) \Sk_{\k_2}(\theta )   \prod_{s=3}^i\sin\te_s\cos\te_{i+1}, \\
 x_N =\Sk_{\k_1}(\tea) \Sk_{\k_2}(\theta )\prod_{s=3}^N\sin\te_s  , 
\end{array}
\label{bb}
\ee
where hereafter a product $\prod_s^i$ such that $s>i$
is assumed to be equal to 1.
The $\k$-dependent trigonometric functions $\Ck_{\k}(x)$ and $\Sk_{\k}(x)$ are 
 defined by \cite{trigo,conf}:
\be 
\Ck_{\k}(x) 
=\left\{
\begin{array}{ll}
  \cos {\sqrt{\k}\, x}, &\   \k >0, \cr 
  1  ,&\  \k  =0 ,\cr 
\cosh {\sqrt{-\k}\, x} ,&\  \k <0 .
\end{array}\right. \qquad 
\Sk_{\k}(x) 
=\left\{
\begin{array}{ll}
    \frac{1}{\sqrt{\k}} \sin {\sqrt{\k}\, x} ,&\   \k >0, \cr 
  x ,&\  \k  =0 ,\cr 
\frac{1}{\sqrt{-\k}} \sinh {\sqrt{-\k}\, x}, &\  \k <0. 
\end{array}\right.  
\label{ae}
\ee
Notice that here $\k\in\{\k_1, \k_2\}$.   The $\k$-tangent is  defined by
$\Tk_\k(x)= {\Sk_\k(x)}/{\Ck_\k(x)}$ and its contraction $\k=0$ gives $\Tk_0(x)= x$.

The   canonical parameters $(\tea,\theta,\te_3,\dots,\te_N)$ dual to   
$(J_{01},J_{12},J_{23},\dots,J_{N-1\,N})$ are called {\em geodesic polar coordinates}.
In order to explain their (physical) geometrical role, let us consider
a  (time-like) geodesic  $l_1$    and other $(N-1)$  (space-like)
geodesics  $l_j$ $(j=2,\dots,N)$  in ${\mathbb S}^N_{[\k_1]\k_2}$ which are orthogonal at
the origin $\cal O$  (each translation  $J_{0i}$ moves $\cal O$ along
$l_i$). Then~\cite{kiev,VulpiLett}:
\begin{itemize}
\itemsep=0pt
\item The radial coordinate $r$ is the   distance between the point $\cal P$ and the origin  $\cal O$
measured along the   geodesic $l$ that joins both points. In the  
Riemannian spaces with
$\k_1=\pm 1/R^2$,
$r$ has dimensions of {\it length}, $[r]=[R]$; notice   that   the 
dimensionless  coordinate $r/R\equiv \te_1$  is usually
taken  instead of $r$, and so considered as an ordinary angle~\cite{17}. In the relativistic spacetimes with
$\k_1=\pm 1/\tau^2$,  $r$ has dimensions of a time-like length:
$[r]=[\tau]$.

\item The coordinate  $\theta$ is an ordinary angle  in the three Riemannian spaces
($\k_2=+1$), say $\theta\equiv\te_2$,  while it
corresponds to a rapidity in the relativistic spacetimes ($\k_2=-1/c^2$)  with dimensions
$[\theta]=[c]$. For the six spaces,  $\theta$   parametrizes the orientation of $l$ with respect
to the basic (time-like) geodesic $l_1$.

\item The remaining $(N-2)$ coordinates $\te_3,\te_4,\dots,\te_N$  are ordinary    angles for
the six spaces and correspond to the  polar angles of $l$ relative
to the reference flag at the origin $\cal O$
 spanned by $\{l_1, l_2\}, \{l_1, l_2,l_3\}, \dots,
\{l_1,\dots,l_{N-1}\}$.

\end{itemize}

In the   Riemannian cases $(\tea,\theta,\te_3,\dots,\te_N)$ parametrize the complete space, while in
the  relativistic spacetimes these only cover the time-like region limited by the light-cone on which
$\theta\to\infty$. The flat contraction $\k_1=0$ gives rise to the usual spherical
coordinates in the Euclidean space ($\k_2=1$).

By introducing  (\ref{bb})  in   (\ref{ce}),   we obtain the
  metric in ${\mathbb
S}^N_{[\k_1]\k_2}$ expressed in geodesic polar coordinates:
\be
 \dd s^2   =\dd\tea^2+\k_2\Sk^2_{\k_1}(\tea) \left\{  \dd\theta^2+
 \Sk^2_{\k_2}(\theta)  \sum_{i=3}^{N} 
\left( \prod_{s={3}}^{i-1} \sin^2 \te_s \right) \dd\te_i^2 \right\} .
 \label{bc}
\ee


\section{Free motion}
 
The metric (\ref{bc}) gives rise to the kinetic energy ${\cal T}$ of a particle  written
in  terms of the velocities $(\dot r,\dot \theta,\dot \te_3,\dots,\dot \te_N)$ which corresponds
to  the free Lagrangian of the geodesic motion on the space ${\mathbb
S}^N_{[\k_1]\k_2}$; namely  
\be
{\cal T}=\frac 12\left(
 \dot\tea^2+\k_2\Sk^2_{\k_1}(\tea) \left\{  \dot\theta^2+
 \Sk^2_{\k_2}(\theta)  \sum_{i=3}^{N} 
\left( \prod_{s={3}}^{i-1} \sin^2 \te_s \right) \dot\te_i^2 \right\} \right).
\label{ea}
\ee
Then the canonical momenta are obtained
through
$p=\partial {\cal T}/\partial \dot q$:
\begin{equation}
\begin{array}{l}
p_r=\dot r,\\
p_{\theta}=\k_2 \Sk_{\k_1}^2(r) \dot\theta,\\[2pt]
p_{\te_i}=\k_2 \Sk_{\k_1}^2(r) \Sk_{\k_2}^2(\theta)\left(  \prod_{s={3}}^{i-1} \sin^2 \te_s
\right)\dot \te_i,
\end{array}
\label{eb}
\end{equation}
where $i=3,\dots,N$. 
Hence
  the free Hamiltonian in the geodesic polar phase space
$(q;p)=(r,\theta,\te_3,\dots,\te_N;p_r,p_{\theta},p_{\te_3},\dots, p_{\te_N})$  with respect
to the canonical Lie--Poisson bracket,
\begin{equation}
\left\{f,g \right\}
=\sum_{i=1}^N\left(\frac{\partial f}{\partial q_i}\frac{\partial g}{\partial p_i}
-\frac{\partial g}{\partial q_i}\frac{\partial f}{\partial p_i}\right),
\label{ec}
\end{equation}
is given by
\begin{equation}
{\cal T}=\frac 12\left(
 p_r^2+\frac 1{\k_2\Sk^2_{\k_1}(\tea) }\left\{  p_{\theta}^2+
\frac 1{ \Sk^2_{\k_2}(\theta) } \sum_{i=3}^{N} 
 \frac{ p_{\te_i}^2 }{\prod_{s={3}}^{i-1} \sin^2 \te_s }\right\} \right).
\label{ed}
\end{equation}

Now we proceed to deduce a   symplectic realization of the Lie generators of
$so_{\k_1,\k_2}(N+1)$. In ambient coordinates
$x_\mu$ and momenta $p_\mu$ this comes from the vector fields (\ref{cf}) through the replacement
$\partial_\mu\to -p_\mu$:
\be 
\begin{array}{l}
J_{01}=x_0 p_1-\k_1 x_1 p_0 ,\qquad
J_{0j}=x_0 p_j-\k_1\k_2 x_j p_0 ,\\
J_{1j}=x_1 p_j-\k_2 x_j p_1 ,\qquad J_{jk}=x_j p_k- x_k p_j , 
\end{array}
\label{eda}
\ee
where $j,k=2,\dots,N$. The metric (\ref{ce}) provides the kinetic energy in the ambient
velocities
$\dot x_\mu$ so that the   momenta $p_\mu$  read
   \begin{equation}
p_0=\dot x_0/\k_1,\quad p_1=\dot x_1,\quad p_j=\k_2\dot x_j,\quad j=2,\dots,N.
\label{edb}
\end{equation}
By computing the velocities $\dot x_\mu$ in the parametrization (\ref{bb}),  
and  introducing the momenta (\ref{eb}) and (\ref{edb}),  we can find   the relationship
between the ambient momenta and the geodesic polar ones (see~\cite{kiev} for $N=3$), which in
turn allows us to   obtain the generators (\ref{eda})  written in   geodesic polar coordinates
and conjugated momenta; these are:

\noindent
$\bullet$ Translation generators ($i=2,\dots,N-1$):
\bea
J_{01}&\!=\!&\Ck_{\k_2}(\theta)p_r-\frac{\Sk_{\k_2}(\theta)}{\Tk_{\k_1}(\tea)}\,p_{\theta},
\nonumber\\
 J_{0i}&\!=\!&\k_2\, \frac{\Sk_{\k_2}(\theta) \prod_{m=3}^{i+1}\sin\te_m  }{\tan\te_{i+1}}\,p_r
+\frac{\Ck_{\k_2}(\theta) \prod_{m=3}^{i+1}\sin\te_m }{\Tk_{\k_1}(\tea)
\tan\te_{i+1}}\,p_{\theta}   \nonumber\\
 &&\quad +\sum_{s=3}^{i+1}\frac{\cos\te_s\prod_{m=3}^{i+1}
\sin\te_m }{\Tk_{\k_1}(\tea)  \Sk_{\k_2}(\theta)\tan\te_{i+1} \prod_{l=3}^{s} \sin\te_l   }\,p_{\te_s}
- \frac{p_{\te_{i+1}} }{\Tk_{\k_1}(\tea) \Sk_{\k_2}(\theta) \prod_{l=3}^{i+1} \sin\te_l     
},\label{trans}\\
J_{0N}&\!=\!&\k_2 \Sk_{\k_2}(\theta) \!\prod_{m=3}^{N}\! \sin\te_m  \,p_\tea    
+\frac{ \Ck_{\k_2}(\theta)\prod_{m=3}^{N} \sin\te_m }{\Tk_{\k_1}(\tea) }\,p_{\theta}
+\sum_{s=3}^{N}\frac{\cos\te_s \prod_{m=s}^{N}
\sin\te_m \, p_{\te_s} }{\Tk_{\k_1}(\tea)    \Sk_{\k_2}(\theta)\prod_{l=3}^{s} \sin\te_l   }.
\nonumber
\eea

\noindent
$\bullet$ Rotation generators ($i=2,\dots,N-1$  and $i<j=3,\dots,N-1$):
\bea
 J_{1i}&\!=\!& \cos\te_{i+1} \prod_{m=3}^{i}\!\sin\te_m  \,p_{\theta}
+\cos\te_{i+1}  \sum_{s=3}^{i}
\frac{   \cos\te_s \prod_{m=s}^{i}  \sin\te_m }
{ \Tk_{\k_2}(\theta) \prod_{l=3}^s\sin\te_l}\,  p_{\te_s} 
- \frac{ \sin\te_{i+1} \, p_{\te_{i+1}}  }{  \Tk_{\k_2}(\theta) \prod_{l=3}^i\sin\te_l  },
\nonumber\\
 J_{1N}&\!=\!& \prod_{m=3}^{N}\!\sin\te_m \, p_{\theta}+
\sum_{s=3}^{N}
\frac{ \cos\te_s \prod_{m=s}^{N}\sin\te_m }
{ \Tk_{\k_2}(\theta) \prod_{l=3}^s\sin\te_l}\,  p_{\te_s},
\nonumber\\
J_{ij}&\!=\!& \sin\te_{i+1}\cos\te_{j+1}\!\prod_{m=i+1}^{j}\!\sin\te_m
\, p_{\te_{i+1}} -\frac{\cos\te_{i+1}\sin\te_{j+1}}
{\prod_{l=i+1}^j\sin\te_l}\, p_{\te_{j+1}} \label{rot}\\
&&\qquad\qquad +\cos\te_{i+1}\cos\te_{j+1}\sum_{s=i+1}^{j}
\frac{ \cos\te_s \prod_{m=s}^{j}\sin\te_m }
{  \prod_{l=i+1}^s\sin\te_l}\,  p_{\te_s} ,\cr
J_{iN}&\!=\!&  \sin\te_{i+1}\!\prod_{m=i+1}^{N}\!\sin\te_m \, p_{\te_{i+1}}
+\cos\te_{i+1}\sum_{s=i+1}^{N}
\frac{ \cos\te_s \prod_{m=s}^{N}\sin\te_m }
{  \prod_{l=i+1}^s\sin\te_l}\,  p_{\te_s}  ,
\nonumber
\eea
where from now on a sum  $\sum_{s=a}^b$ such that $a>b$
is assumed to be equal to 0.

Then  the following statement holds.
\medskip

\noindent
{\bf Proposition 1.}
{\it (i) The generators (\ref{trans}) and (\ref{rot}) fulfil the commutation relations 
(\ref{aax}) with respect to the  Lie--Poisson bracket (\ref{ec}).\\
(ii) All of them  Poisson
commute with
$\cal T$ (\ref{ed}).}
 \medskip

 The first point can be proven by direct computations, while the second one comes from the fact
that
 the kinetic energy can be obtained 
from  the Casimir ${\cal C}$ (\ref{bba}) as $2\k_2{\cal T}={\cal C}$ by introducing the
above realization of the generators.

Therefore all the $N(N+1)/2$ generators of ${so}_{\k_1,\k_2}(N+1)$ give rise to integrals of  
motion for
$\cal T$.  In order to characterize the maximal superintegrability of the geodesic motion on the 
space  ${\mathbb S}^N_{[\k_1]\k_2}$ let us define  two sets of $(N-1)$ functions, coming from
the rotation generators,  which are quadratic in the momenta:
\begin{equation}
\begin{array}{l}
\displaystyle{ {\cal J}^{(l)} =\sum_{j=2}^l J_{1j}^2 +\k_2\sum_{i,j=2}^l J_{ij}^2 ,\quad 
l=2,\dots,N,}\\
\displaystyle{     {\cal J}_{(N)}= {\cal J}^{(N)},
\quad {\cal J}_{(k)} =\!\!\sum_{i,j=N-k+1}^N\!\!
J_{ij}^2     ,\quad k=2,\dots,  N-1.}
\end{array}
\label{integralsa}
\end{equation}
And it can be shown:

\medskip

\noindent
{\bf Proposition 2.} 
{\it 
(i) The   $N$  functions $\{
{\cal J}^{(2)},{\cal J}^{(3)},\dots,{\cal J}^{(N)},\cal T\} $ are mutually   in
involution.
The same property holds for the second set    $\{
{\cal J}_{(N)},{\cal J}_{(N-1)},\dots,{\cal J}_{(2)},\cal T\} $.

\noindent
(ii) The $2N-1$ functions $\{
{\cal J}^{(2)}, {\cal J}^{(3)},\dots,
 {\cal J}^{(N)}\equiv  {\cal J}_{(N)},\dots,{\cal J}_{(3)}, 
{\cal J}_{(2)}, J_{0j},\cal T\}$, where $j$ is fixed ($j=1,\dots,N$), are functionally
independent. 
}

\medskip

 Consequently,  $\cal T$ is MS and its independent integrals of motion
come from the (Lorentz) rotation subgroup plus  one from the 
translation generators.


\section{Quasi-maximally superintegrable potentials}

Let us consider the following potential   ${\cal U}(q)={\cal U}(r,\theta,\te_3,\dots,\te_N)$ defined
on   ${\mathbb S}^N_{[\k_1]\k_2}$:
\begin{eqnarray}
  {\cal U}&\!=\!&{\cal
F}'(x_0)+  \sum_{s=1}^N\frac{\beta_i}{x_i^2}\label{fa}\\
&\!=\!& {\cal F}(r)
+\frac{1}{\Sk_{\k_1}^2(r)}\left(\frac{\beta_1}{\Ck_{\k_2}^2(\theta)}+
\sum_{i=2}^{N-1}\frac{\beta_i}{\Sk_{\k_2}^2(\theta) \prod_{s=3}^i\sin^2\te_s\cos^2\te_{i+1}}
+\frac{\beta_N}{\Sk_{\k_2}^2(\theta)\prod_{s=3}^N\sin^2\te_s  } \right),
\nonumber
\end{eqnarray}
where ${\cal F}'(\Ck_{\k_1}(r))\equiv {\cal
F}(r)$ is an arbitrary smooth function and $\beta_i$ are
arbitrary real constants. This corresponds to the superposition of a (curved) central potential
${\cal F}(r)$,  only depending on the geodesic distance $r$,   with $N$ centrifugal barriers
associated with the $\beta_i$-terms. Therefore this is the generalization of the
Euclidean potential (\ref{aa}) to ${\mathbb S}^N_{[\k_1]\k_2}$.

 The   Hamiltonian ${\cal H}={\cal T}+ {\cal U}$, with kinetic
energy (\ref{ed}) and potential (\ref{fa}), has $N(N-1)/2$ integrals of the motion
quadratic in the momenta which come from  the  rotation generators (\ref{rot}) 
 $(i,j=2,\dots,N)$:
\begin{equation}
 I_{1i}=J_{1i}^2+2\beta_1\k_2^2\,\frac{x_i^2}{x_1^2}+2\beta_i\k_2 
\,\frac{x_1^2}{x_i^2} ,\quad I_{ij}=J_{ij}^2+2\beta_i\k_2\,\frac{x_j^2}{x_i^2}+2\beta_j\k_2
\,\frac{x_i^2}{x_j^2} .
\label{int}
\end{equation}
In the geodesic polar phase space these are given by $(i,j=2,\dots,N-1)$:
\begin{equation}
\begin{array}{l}
\displaystyle{I_{1i}= J_{1i}^2+2\beta_1\k_2^2\Tk_{\k_2}^2(\theta)   \prod_{s=3}^i\sin^2\te_s   
\cos^2\te_{i+1} +\frac{2\beta_i\k_2}{ \Tk_{\k_2}^2(\theta)   \prod_{s=3}^i\sin^2\te_s   
\cos^2\te_{i+1} }},\\ 
\displaystyle{I_{1N}= J_{1N}^2+2\beta_1\k_2^2\Tk_{\k_2}^2(\theta )   \prod_{s=3}^N\sin^2\te_s   
  +\frac{2\beta_N\k_2}{ \Tk_{\k_2}^2(\theta )   \prod_{s=3}^N\sin^2\te_s     }},\\ 
\displaystyle{I_{ij}= J_{ij}^2+2\beta_i\k_2 \frac{\prod_{s=i+1}^j\sin^2\te_s \cos^2\te_{j+1}
}{\cos^2\te_{i+1}} +  \frac{ 2\beta_j\k_2\cos^2\te_{i+1}} {\prod_{s=i+1}^j\sin^2\te_s \cos^2\te_{j+1}
} },\\ 
\displaystyle{I_{iN}= J_{iN}^2+2\beta_i\k_2 \frac{\prod_{s=i+1}^N\sin^2\te_s  
}{\cos^2\te_{i+1}} +  \frac{ 2\beta_N\k_2\cos^2\te_{i+1}} {\prod_{s=i+1}^N\sin^2\te_s  }.}
\end{array}
\label{fc}
\end{equation}

Obviously, neither    all these constants are in involution, nor they are functionally
independent.  Similarly to (\ref{integralsa}), we   define  two sets of $(N-1)$ functions:
\begin{equation}
\begin{array}{l}
\displaystyle{ {  Q}^{(l)} =\sum_{j=2}^l I_{1j}  +\k_2\sum_{i,j=2}^l I_{ij} ,\quad 
l=2,\dots,N,}\\
\displaystyle{     {  Q}_{(N)}= {  Q}^{(N)},
\quad {  Q}_{(k)} =\!\!\sum_{i,j=N-k+1}^N\!\!
I_{ij}     ,\quad k=2,\dots,  N-1.}
\end{array}
\label{integralsb}
\end{equation}
And superintegrability properties of ${\cal H}$ are determined by:
\medskip

\noindent
{\bf Proposition 3.} 
{\it 
(i) The   $N$  functions $\{
{Q}^{(2)},{Q}^{(3)},\dots,{Q}^{(N)},\cal H\} $ are mutually   in
involution.
The same   holds for the   set    $\{
{Q}_{(N)},{Q}_{(N-1)},\dots,{Q}_{(2)},\cal H\} $.\\
\noindent
(ii) The $2N-2$ functions $\{
{Q}^{(2)}, {Q}^{(3)},\dots,
 {Q}^{(N)}\equiv  {Q}_{(N)},\dots,{Q}_{(3)}, 
{Q}_{(2)},\cal H\}$ are functionally
independent. 
}
\medskip

 Notice that the difference with respect to the free motion described in proposition 2 is that now 
{\it one} constant of the motion is left to ensure maximal superintegrability (for ${\cal T}$ this
role was played by  one of the translations generators). In this sense we  shall say that  ${\cal
H}$ is   {\it quasi-maximally superintegrable}. Nevertheless, some specific choices for the
arbitrary radial function ${\cal F}(r)$ lead to  an additional integral thus providing
MS potentials. Next we present the two relevant cases
which correspond to the  SW   and the generalized KC
systems on the space
${\mathbb S}^N_{[\k_1]\k_2}$.


\section{Maximally superintegrable potentials}


\subsection{Smorodinsky--Winternitz potential}

The harmonic oscillator potential on  
 ${\mathbb S}^N_{[\k_1]\k_2}$ is obtained through   the following   choice for the   function
$\cal F$:
\begin{equation}
{\cal
F}'(x_0)=\beta_0\left(\frac{1-x_0^2}{\k_1 x_0^2} \right)=
\beta_0\left(\frac{ x_1^2+ \k_2 \sum_{i=2}^N x_i^2   }{x_0^2} \right),\quad
{\cal F}(r)=\beta_0\Tk^2_{\k_1}(r),
\label{ga}
\end{equation}
where $\beta_0$ is an arbitrary real parameter $(\beta_0=\omega^2)$. This is just the Higgs
oscillator~\cite{Higgs,Leemon} formerly obtained in the curved Riemannian spaces. 
We can add the $N$ arbitrary centrifugal terms (\ref{fa}) to (\ref{ga}) thus obtaining the
generalization of the SW  system, ${\cal H}^{\rm
SW}={\cal T}+{\cal U}^{\rm SW}$, to the space  
${\mathbb S}^N_{[\k_1]\k_2}$:
 \begin{equation}
\begin{array}{l}
\displaystyle{ {\cal U}^{\rm SW}=   \beta_0\Tk^2_{\k_1}(r) 
+\frac{1}{\Sk_{\k_1}^2(r)}\left(\frac{\beta_1}{\Ck_{\k_2}^2(\theta)}  \right.   } \\ 
\displaystyle{ \qquad\qquad +\left.
\sum_{i=2}^{N-1}\frac{\beta_i}{\Sk_{\k_2}^2(\theta) \prod_{s=3}^i\sin^2\te_s\cos^2\te_{i+1}}
+\frac{\beta_N}{\Sk_{\k_2}^2(\theta)\prod_{s=3}^N\sin^2\te_s  } \right).}
\end{array}
\label{gb}
\end{equation}
   The contraction $\k_1=0$ (with $\k_2=+1$) of (\ref{gb})
reproduces the flat SW potential given in the Introduction but here written in polar
coordinates. Notice that, under this contraction, the $N$ ambient coordinates $x_i$
$(i=1,\dots,N)$ coincide with the Cartesian ones, while
$x_0=1$ (see (\ref{bb})). The 2D and 3D
SW systems on the   spherical and hyperbolic spaces   have been constructed  by following
different approaches~\cite{11,18,20,21,27}, and for the three $N$D Riemannian spaces altogether
these can be found in~\cite{VulpiLett,CRMVulpi,angellett}. Less developed are the SW
Hamiltonians on relativistic spacetimes  since, to our knowledge, only very recent results
cover the (1+1)D~\cite{jpa2D,car2} and (2+1)D cases~\cite{kiev}. Moreover,   SW-type systems on
certain 2D~\cite{jpa2D} and
$N$D~\cite{sigmaorlando,enciso} spaces of {\em nonconstant} curvature have been, again very
recently, studied.

 The SW Hamiltonian on   ${\mathbb S}^N_{[\k_1]\k_2}$ has additional constants of motion to those
given in proposition 3. Similarly to what happened with the geodesic motion,   any of the translation
generators   (\ref{trans})  gives rise to an integral   quadratic in
the momenta 
 $(i=2,\dots,N)$:
\begin{equation}
I_{01}=J_{01}^2+2\beta_0\,\frac{x_1^2}{x_0^2}+2\beta_1\,
\frac{x_0^2}{x_1^2} ,\quad I_{0i}=J_{0i}^2+2\beta_0\k_2^2\,\frac{x_i^2}{x_0^2}+2\beta_i\k_2
\,\frac{x_0^2}{x_i^2} , 
\label{inta}
\end{equation}
to be compared with (\ref{int}). In polar coordinates, these are $(i=2,\dots,N-1)$:
\bea
 I_{01}&\!=\!&J_{01}^2+2\beta_0 \Tk_{\k_1}^2(r)\Ck_{\k_2}^2(\theta)+
 \frac{2\beta_1}{\Tk_{\k_1}^2(r)\Ck_{\k_2}^2(\theta)}  ,\nonumber\\
 I_{0i}&\!=\!&J_{0i}^2+2\beta_0\k_2^2
\Tk_{\k_1}^2(r)\Sk_{\k_2}^2(\theta)       \prod_{s=3}^i\sin^2\te_s   
\cos^2\te_{i+1}     +
\frac{2\beta_i\k_2}{   \Tk_{\k_1}^2(r)\Sk_{\k_2}^2(\theta)       \prod_{s=3}^i\sin^2\te_s   
\cos^2\te_{i+1}     }  ,\nonumber\\ 
 I_{0N}&\!=\!&J_{0N}^2+2\beta_0\k_2^2
\Tk_{\k_1}^2(r)\Sk_{\k_2}^2(\theta)       \prod_{s=3}^N\sin^2\te_s    +
\frac{2\beta_N\k_2}{   \Tk_{\k_1}^2(r)\Sk_{\k_2}^2(\theta)       \prod_{s=3}^N\sin^2\te_s }.
\label{gd}
\eea

From this set of $N$ additional integrals, we establish the superintegrability of ${\cal H}^{\rm SW}$.
\medskip

\noindent
{\bf Proposition 4.}
{\it (i) The $N$ functions (\ref{gd})   Poisson commute with   ${\cal
H}^{\rm SW}$.\\
(ii) The $2N-1$ functions $\{
{Q}^{(2)}, {Q}^{(3)},\dots,
 {Q}^{(N)}\equiv  {Q}_{(N)},\dots,{Q}_{(3)}, 
{Q}_{(2)},I_{0j},\cal H\}$, where
$j$ is fixed $(j=1,\dots,N)$,  are functionally
independent. }
\medskip

Therefore, the   known result concerning maximal superintegrability of the SW system on the
three
$N$D Riemannian spaces of constant curvature  also holds for the
relativistic spacetimes covering in a unified way the complete family ${\mathbb S}^N_{[\k_1]\k_2}$.


\subsection{Generalized Kepler--Coulomb potential}

The    KC  potential~\cite{car1,car2,18,21,
Schrodingerdual,Schrodingerdualc,Schrodingerdualb,27,Schrodinger} on the space ${\mathbb
S}^N_{[\k_1]\k_2}$ is achieved by choosing
\begin{equation}
{\cal
F}'(x_0)=-k \,\frac{x_0}{\sqrt{(1-x_0^2)/\k_1}} =
-k\, \frac{x_0}{\sqrt{x_1^2+\k_2 \sum_{j=2}^N x_j^2  }},\quad
{\cal F}(r)=-\frac{k }{\Tk_{\k_1}(r)},
\label{ha}
\end{equation}
where $k$ is an arbitrary real parameter. Such a potential is known to be  MS
on the three $N$D Riemannian spaces. Nevertheless, in this case it is not possible to add $N$
arbitrary centrifugal terms keeping this property as it does happen with the SW potential in
such a manner that, at least,    one of the $\beta_i$-terms must vanishes  (see~\cite{kiev}
for the 3D case).  In this way,   we find that, in principle, there are $N$ possible
generalized KC (GKC) potentials, which can be  understood as the superposition of the proper 
KC potential (\ref{ha}) together with $(N-1)$ centrifugal terms appearing within
(\ref{fa}). Explicitly, these are $(j=2,\dots,N-1)$:
\bea
 {\cal U}^{\rm
GKC}_1&\!=\!& -\frac{k
}{\Tk_{\k_1}(r)} 
+\frac{1}{\Sk_{\k_1}^2(r)\Sk_{\k_2}^2(\theta)}\left( 
\sum_{l=2}^{N-1}\frac{\beta_l}{ \prod_{s=3}^l\sin^2\te_s\cos^2\te_{l+1}}
  +\frac{\beta_N}{\prod_{s=3}^N\sin^2\te_s  } \right) , \nonumber \\ 
 {\cal U}^{\rm
GKC}_j&\!=\!&-\frac{k
}{\Tk_{\k_1}(r)} 
+\frac{1}{\Sk_{\k_1}^2(r)}\left(\frac{\beta_1}{\Ck_{\k_2}^2(\theta)}
\phantom{\frac{\beta_N}{ \prod_{s=3}^N  }} \right.    \nonumber   \\ 
 & & \qquad\quad   \left. +
\sum_{l=2; l\ne j}^{N-1}\frac{\beta_l}{\Sk_{\k_2}^2(\theta)
\prod_{s=3}^l\sin^2\te_s\cos^2\te_{l+1}}
+\frac{\beta_N}{\Sk_{\k_2}^2(\theta)\prod_{s=3}^N\sin^2\te_s  } \right) , \label{hb}
 \\ 
 {\cal U}^{\rm
GKC}_N&\!=\!&-\frac{k
}{\Tk_{\k_1}(r)} 
+\frac{1}{\Sk_{\k_1}^2(r)}\left(\frac{\beta_1}{\Ck_{\k_2}^2(\theta)}+
\sum_{l=2}^{N-1}\frac{\beta_l}{\Sk_{\k_2}^2(\theta) \prod_{s=3}^l\sin^2\te_s\cos^2\te_{l+1}}
 \right). 
\nonumber
\eea
The contraction of a given ${\cal U}^{\rm
GKC}_i$ to the Euclidean case gives the known result~\cite{6,Miguel,Williams} ${\cal U}^{\rm
GKC}_{i}= - {k}/{r}  + \sum_{l=1;l\ne i}^N  {\beta_l}/{x_l^2} $ as commented in the Introduction.

For 
  each of the $N$ potentials   ${\cal U}^{\rm
GKC}_i$  there exists an additional constant of the motion
given by $(i=1,\dots,N)$:
\begin{equation}
 L_i=\sum_{l=1; l\ne i}^N J_{0l}J_{li}+k\,\frac{\k_2
x_i}{\sqrt{x_1^2+\k_2 \sum_{j=2}^N x_j^2  }}-2\k_2\sum_{l=1; l\ne i}^N \beta_l\,\frac{x_0x_i}{x_l^2} ,
\label{intb}
\end{equation}
where   $J_{li}=-J_{il}$ if $i<l$. In    the geodesic polar phase space these
integrals turn out to be $(j=2,\dots,N-1)$:
\bea
 L_1&\!=\!& -\sum_{l=2 }^N J_{0l}J_{1l}+k\,\k_2 \Ck_{\k_2}(\theta)
 \nonumber\\ 
 & &   
-\frac{2\k_2\Ck_{\k_2}(\theta)}{\Tk_{\k_1}(r)\Sk_{\k_2}^2(\theta)}
\left(\sum_{l=2}^{N-1} \frac{\beta_l}{\prod_{m=3}^l\sin^2\te_m  \cos^2\te_{l+1} }
+\frac{\beta_N}{\prod_{m=3}^N\sin^2\te_m} \right),\nonumber\\ 
 L_j&\!=\!& \sum_{l=1;l\ne j }^N J_{0l}J_{lj}+k\,\k_2 \Sk_{\k_2}(\theta)
\prod_{s=3}^j\sin\te_s  \cos\te_{j+1} 
-2\k_2  \frac{  \prod_{s=3}^j\sin\te_s\cos\te_{j+1}  }{\Tk_{\k_1}(r)\Sk_{\k_2}(\theta)}
\left(\beta_1 \Tk_{\k_2}^2(\theta) \phantom { \frac{\beta_N} {\prod_{i}^N} }\right.
  \nonumber\\ 
   & & \qquad\qquad \qquad\qquad \qquad \qquad  \left.  +
\sum_{l=2;l\ne j}^{N-1} \frac{\beta_l}{\prod_{m=3}^l\sin^2\te_m  \cos^2\te_{l+1} }
+\frac{\beta_N}{\prod_{m=3}^N\sin^2\te_m} \right), \label{hc}\\ 
 L_N &\!=\!&  \sum_{l=1  }^{N-1} J_{0l}J_{lN}+k\,\k_2 \Sk_{\k_2}(\theta)
\prod_{s=3}^N\sin\te_s  
  \nonumber\\ 
 & &  
-2\k_2\, \frac{  \prod_{s=3}^N\sin\te_s  }{\Tk_{\k_1}(r)\Sk_{\k_2}(\theta)}
\left(\beta_1 \Tk_{\k_2}^2(\theta)+
\sum_{l=2 }^{N-1} \frac{\beta_l}{\prod_{m=3}^l\sin^2\te_m  \cos^2\te_{l+1} }
  \right)  .  
\nonumber
\eea
 The  MS of   each    Hamiltonian ${\cal H}^{\rm GKC}_i={\cal T}+{\cal
U}^{\rm GKC}_i$ ($i$ fixed and  $i=1,\dots,N $)  is stated as:
\medskip

\noindent
{\bf Proposition 5.}
{\it (i) The function $L_{i}$ (\ref{hc}) Poisson commutes with   ${\cal
H}^{\rm GKC}_i$.\\
(ii) The $2N-1$ functions $\{
{Q}^{(2)}, {Q}^{(3)},\dots,
 {Q}^{(N)}\equiv  {Q}_{(N)},\dots,{Q}_{(3)}, 
{Q}_{(2)},L_i,{\cal
H}^{\rm GKC}_i\}$   are functionally
independent. }
\medskip

We remark that for the three Riemannian cases with $\k_2=+1$, the $N$ GKC Hamiltonians are
all equivalent providing the superposition of the KC potential with $(N-1)$
centrifugal barriers. In contrast, for the three relativistic spacetimes with 
$\k_2<0$, ${\cal U}^{\rm GKC}_1$ is formed by a time-like KC potential  with
$(N-1)$  space-like centrifugal barriers, while the remaining $(N-1)$ potentials ${\cal U}^{\rm
GKC}_j$ $(j=2,\dots,N)$ are all equivalent and composed by the time-like KC potential, a
time-like centrifugal barrier with parameter
$\beta_1$, and other $(N-2)$  space-like ones. In any case, to consider   initially 
$N$  possible GKC Hamiltonians affords for a direct  understanding of the  appearance of the
Laplace--Runge--Lenz vector on 
${\mathbb S}^N_{[\k_1]\k_2}$ as the following statements show.

\medskip

\noindent
{\bf Proposition 6.}
{\it   Let us take the Hamiltonian ${\cal H}^{\rm GKC}_i={\cal
T}+{\cal U}^{\rm GKC}_i$ ($i$ fixed and $i=1,\dots,N$) with $\beta_j = 0$ $(j\ne i)$. Then\\
(i) The  two functions $L_i,L_{j}$ Poisson commute with   ${\cal
H}^{\rm GKC}_i$.\\
(ii) The set $\{
{Q}^{(2)}, {Q}^{(3)},\dots,
 {Q}^{(N)}\equiv  {Q}_{(N)},\dots,{Q}_{(3)}, 
{Q}_{(2)}, {\cal
H}^{\rm GKC}_i\}$    together with
either $L_i$ or $L_j$ are  $2N-1$ functionally independent functions. }
\medskip

\noindent
{\bf Proposition 7.}
{\it Let  $\beta_i=0$ $\forall i$, then: \\ 
(i) The $N$ GKC potentials reduce to its common KC potential on ${\mathbb
S}^N_{[\k_1]\k_2}$:  ${\cal U}^{\rm GKC}_i\equiv
{\cal U}^{\rm KC}=-k/\Tk_{\k_1}(r)$.\\
(ii) The $N$
functions  $(j=2,\dots,N-1)$:
\bea
 L_1&\!=\!& -\sum_{l=2 }^N J_{0l}J_{1l}+k\,\k_2 \Ck_{\k_2}(\theta) ,\nonumber\\ 
 L_j&\!=\!& \sum_{l=1;l\ne j }^N J_{0l}J_{lj}+k\,\k_2 \Sk_{\k_2}(\theta)
\prod_{s=3}^j\sin\te_s  \cos\te_{j+1}  , \label{hhc}\\ 
 L_N &\!=\!&  \sum_{l=1  }^{N-1} J_{0l}J_{lN}+k\,\k_2 \Sk_{\k_2}(\theta)
\prod_{s=3}^N\sin\te_s  , 
\nonumber
\eea
Poisson commute  with   ${\cal H}^{\rm KC}={\cal
T}+{\cal U}^{\rm KC}$. \\
(iii) The  set $\{
{Q}^{(2)}, {Q}^{(3)},\dots,
 {Q}^{(N)}\equiv  {Q}_{(N)},\dots,{Q}_{(3)}, 
{Q}_{(2)}, {\cal H}^{\rm KC} \}$    together with
  any of the components $L_i$ ($i=1,\dots,N$) are  $2N-1$ functionally independent functions.}
\medskip

We stress that  (\ref{hhc}) are the components of the Laplace--Runge--Lenz $N$-vector
on  ${\mathbb S}^N_{[\k_1]\k_2}$; these are transformed as a vector under the action of
the generators of the subgroup $SO_{\k_2}(N)$ (\ref{bca}) (either rotations for $\k_2>0$ or Lorentz
transformations for $\k_2<0$).

Proofs and details of all the results here presented will be given elsewhere, together with a 
  physical/geometrical description of the MS SW and GKC Hamiltonians 
on each particular space ${\mathbb S}^N_{[\k_1]\k_2}$.


\section*{ACKNOWLEDGEMENTS}

 This work was partially supported  by the Ministerio de Educaci\'on y
Ciencia   (Spain, Project FIS2004-07913) and  by the Junta de Castilla y
Le\'on   (Spain, Project  VA013C05).


\end{document}